\documentclass[nohyper,notoc]{JHEP} 
\usepackage{epsfig}

\def \lsim{\mathrel{\vcenter
     {\hbox{$<$}\nointerlineskip\hbox{$\sim$}}}}

\def\beq{\begin{equation}}
\def\eeq{\end{equation}}
\def\bea{\begin{eqnarray}}
\def\eea{\end{eqnarray}}
\def\bq{\begin{quote}}
\def\eq{\end{quote}}

\preprint{CERN-TH/99-384}

\title{Updated Bounds on Milli-Charged Particles}

\author{Sacha Davidson\\
Theoretical Physics, Oxford University, 1 Keble Road, Oxford OX1 3NP, 
Great Britain} 

\author{Steen Hannestad\\
NORDITA, Blegdamsvej 17, 2100 Copenhagen, Denmark}

\author{Georg Raffelt\\
Max-Planck-Institut f\"ur Physik 
(Werner-Heisenberg-Institut), 
F\"ohringer Ring 6, 80805 M\"unchen, Germany}

\abstract{We update the bounds on fermions with electric charge
  $\epsilon e$ and mass $m_\epsilon$.  For $m_\epsilon\lsim m_e$ we
  find $10^{-15}\lsim\epsilon<1$ is excluded by laboratory
  experiments, astrophysics and cosmology. For larger masses, the
  limits are less restrictive and depend on $m_\epsilon$. For
  milli-charged neutrinos, the limits are stronger,
  especially if the different flavors mix as suggested by current
  experimental evidence.}


\begin{document} 


\section{Introduction}

A puzzle of continuing
interest in particle physics is
the apparent quantization of electric charge.
The electric charge of all the particles we
see appears \cite{MM}
to be an integer multiple of $Q_e/3$,
where $Q_e$ is the electron charge.
However it is theoretically
consistent to have  particles 
with electric  charge $\epsilon Q_e$
where $\epsilon$ is any real number. We
consider $\epsilon < 1$ in this paper,
and refer to these particles as
``milli-charged.''

Milli-charged particles can be
introduced into the Standard Model
in a variety of ways. One is  to add
an $SU(3) \times SU(2)$ singlet
Dirac fermion with hypercharge
$Y = 2 \epsilon$.  However, this is 
not simple \cite{OVZ} if the hypercharge
$U(1)_Y$ is embedded in a grand unified 
gauge group. A second way of
generating effectively milli-charged particles,
which works even if hypercharge is quantised,
is due to Holdom \cite{Bob1}.
He introduced a second unbroken
 ``mirror'' $U(1)'$, and showed that the photon
and the ``paraphoton'' can mix,
so that a particle charged under the
$U(1)'$ appears to have a small coupling
to the photon. A third mechanism
for introducing milli-charged
particles is to remain with the
Standard Model particle content and
allow the neutrinos to have small
electric charges \cite{Foot}.
If the Standard Model hypercharge
operator $Y_{SM}$ is redefined to be
$Y' = Y_{SM} + 2 \sum_i \epsilon_i (B/3 - L_i)$,
neutrinos acquire small electric charges
$\epsilon_i$  and the Standard Model
anomaly cancellation is preserved.
We assume here that $U(1)_{em}$ is
an unbroken symmetry; it can be
spontaneously broken in models
with milli-charged scalars \cite{IKS},
but the bounds in this case are
slightly different from what
we discuss here \cite{TT} (for
instance the photon has a mass).

The purpose of this paper is to
collect and update  experimental and
observational bounds on milli-charged
particles. Such limits have been previously
considered for  milli-charged
fermions that are not neutrinos  \cite{Bob2,G+H,G+R,D+I,M+R,B+H,DCB,M+N,D+P}
and for milli-charged neutrinos
\cite{neu,FJL,TT1,B+V,FL}. See also \cite{GGR} for a discussion
of astrophysical constraints on
all types of milli-charged particles.
We recalculate the
laboratory bounds 
using recent data, and include
the limits from recent experiments
searching for milli-charged particles \cite{japan,SLAC}.
We numerically evaluate the
constraint from Big Bang Nucleosynthesis
on particles with milli-hypercharge,
and find a bound similar to previous 
estimates. We also revisit the astrophysical
constraints, resolving a discrepancy in
the White Dwarf bounds between \cite{D+I}
and \cite{DCB} (in favour of the
stronger bound of \cite{D+I}). Our
Supernova limit differs slightly
from what is found in Ref.~\cite{M+R}, as will be explained
later, and the
Red Giant bound remains unchanged.
We quote the  previous estimated bounds
\cite{DCB}
from balloon
experiments searching for strongly 
interacting Dark Matter, and from underground WIMP detectors.
For a more detailed discussion of
the calculations, see for instance \cite{DCB,D+P,B+V,GGR}.

Our conclusions are presented in Fig.~1
which is an update of 
Fig.~1 of Ref.~\cite{D+P}.  It presents
bounds on milli-charged fermions in models
with  an extra $U(1)'$, and without.  Milli-charged
neutrinos will be briefly discussed at the end of
the paper.  
 The limits from laboratory
experiments, from Red
Giants and  White Dwarfs are independent of
whether there is a paraphoton,
and appear as solid lines. We discuss the
Supernova and nucleosynthesis bounds with
some care for the model without
a paraphoton; the limits should be similar
in the presence of a paraphoton \cite{D+P} so
these bounds also appear as solid lines.  The limit
from  $\Omega < 1$ differs in the two models: in
the absence of a paraphoton, the millicharge
annihilation cross-section is proportional to
$\epsilon^2/m^2$. So the excluded area, indicated
with dotted lines, extends to lower
masses for smaller $\epsilon$. The bound
on models with  a paraphoton is the horizontal dashed line.

\begin{figure}[t]
\begin{center}
\epsfxsize=14truecm\epsfbox{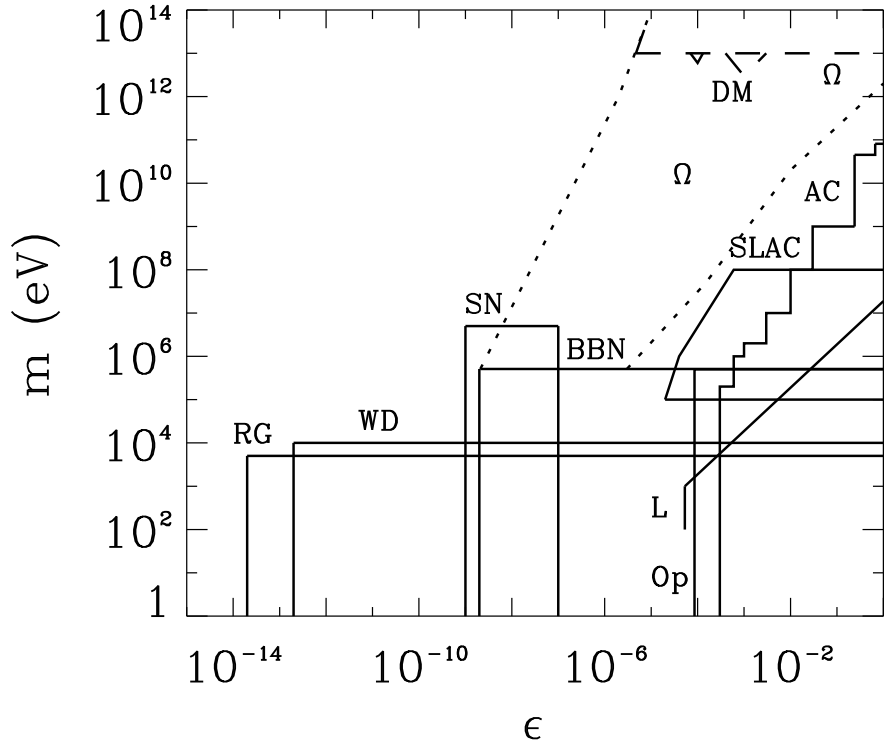}
\vspace{0truecm}
\end{center}
\caption{Regions of mass-charge space ruled out for milli-charged
particles.   The solid and dashed
lines apply to the model with a paraphoton;
solid and dotted lines apply in the absence of 
a paraphoton. The bounds arise from
the following constraints: AC---accelerator experiments; 
Op---the Tokyo search for  the invisible
decay of ortho-positronium \protect\cite{japan};
SLAC---the SLAC milli-charged particle
search \protect\cite{SLAC}; L---the Lamb
shift; BBN---nucleosynthesis; 
$\Omega$ ---$\Omega < 1$; RG---plasmon decay in
red giants; WD---plasmon decay in white dwarfs; DM---dark 
matter searches;
SN---Supernova 1987A.}
\label{fig1}
\end{figure}


\section{Laboratory Bounds}

We plot  four lines  on Figure 1 from
laboratory data: 1) a combined ``accelerator''
line consisting of the limits from LEP
and beam dump experiments \cite{G+H,DCB}\footnote{
The values of $m_\epsilon$ excluded by direct
searches  are quoted in Table II of \cite{DCB}, but the values
given are in MeV, not GeV as claimed in the caption.  
The limit is placed correctly,
however, in Fig. 2 of \cite{DCB}.},
2) the bound found by the Tokyo group \cite{japan}
from the non-observation of invisible Ortho-positronium decay,
3) the 
limit from the dedicated milli-charged
particle search experiment at SLAC \cite{SLAC}, and
4) an updated constraint from the Lamb shift
using more recent data \cite{Lamb}. 

LEP has taken many years of data since limits
from LEP were previously considered \cite{DCB},
so we briefly discuss possible bounds,
although these are  weak
because the milli-charge coupling
to the $Z$ is suppressed by $\sin^2 \theta_W$.
A search for
particles with fractional charge $\epsilon = 2/3$
was performed  by OPAL
using 1991-93 data \cite{LEP1},  which rules
out $\epsilon \geq 2/3$ for $m_{\epsilon} < 84$ GeV.
This
bound could be extended to the present
kinematic limit $m_{\epsilon} < 100$ GeV,
if one assumes that  a particle with $1> \epsilon > 2/3$
would be seen as such in the detector. 
Fractionally charged particles would
contribute to the invisible width of the
$Z$  if they were not seen in the detector
\footnote{This bound assumes that particles
with 1/4 $ < \epsilon < 2/3$ would
look like noise in the detector, and
not be mis-identified as $\epsilon = 1$ tracks.}, in which case 
the LEP bound  can be extended 
to 
\beq
\epsilon < 0.24 ~~~~m_{\epsilon} > 45 {\rm ~ GeV},
\eeq
from requiring that
that milli-charges not contribute
more than the $ 2 \sigma$ error
to the invisible width of the $Z$ at LEP1.

We calculate an improved bound from  the Lamb shift
measurements  by requiring
that the milli-charged particle  
vacuum polarisation contribution
 be less than the  $2\sigma$ error ( = 20 $ \times 10^{-3} $ MHz) \cite{Lamb}.
This is a more
recent measurements of the $2S_{1/2}-2P_{1/2}$  Lamb shift
\cite{Lamb} than used in \cite{DCB}.
There are  more precise measurements
of combinations of Lamb shifts  of
different principle quantum number $n$ ($n \neq 2$) \cite{Lamb2};
however they do not
substantially improve our bound so
we use the  $2S_{1/2}-2P_{1/2}$ results
because the milli-charged contribution is
easier to disentangle.
The bound from the Lamb shift is stronger than
from $g-2$, despite the fact that $g-2$ is
measured very precisely \cite{g2}, because
milli-charged particles contribute to the
Lamb shift at one loop
but to $g-2$ at two loop.
For a discussion of
the effects of milli-charged particles
on precision QED measurements, see \cite{D+I,AC}.


\section{Big-Bang Nucleosynthesis}

\subsection{Dominant Processes}

Particles with small electric charge will interact with the plasma in
the early universe and in this way get thermally excited. Therefore,
their primordial energy density and thus their interaction strength
can be constrained by the standard nucleosynthesis limit on $N_{\rm eff}$, the
effective number of thermally excited neutrino degrees of freedom. A
milli-charged fermion must be a Dirac particle and thus has four
intrinsic degrees of freedom. If it were fully excited in the early
universe, it would contribute $\Delta N_{\rm eff}=2$
prior to the annihilation of electrons and positrons
\footnote{This is when the BBN bound on the number of
neutrinos applies}.
Such a particle would be excluded by a large margin.
 During this
annihilation process entropy is transferred to the milli-charged
particles, heating them relative to the neutrinos. 
At the same time, the photon temperature, $T_\gamma$, relative
to the neutrino temperature, $T_\nu$, is less than in the standard
model. Normally, $T_\nu/T_\gamma \simeq 0.714$, whereas here
$T_\nu/T_\gamma \simeq 0.849$. Altogether, neutrinos plus milli-charged
particles contribute $N_{\rm eff}=3 \times (0.849/0.714)^{4} +
2 \times 1.401^4 = 13.7$, so that 
$\Delta N_{\rm eff}=10.7$ at late times.

While it is easy to estimate  $\Delta N_{\rm eff}$ for
milli-charged particles
where full thermal equilibrium obtains, we can derive a
more restrictive limit by studying its value as a function
of $\epsilon$ in a regime where full equilibrium is not achieved.  To
this end we assume that the new particles interact only
electromagnetically by their milli-charge.
The bounds in the case where there is an extra
$U(1)$ are at least as restrictive, due to the presence
of the paraphoton \cite{D+P}.

We find that the dominant electromagnetic processes which couple the
gas of milli-charged particles $f$ to the electromagnetic plasma are
\begin{eqnarray}\label{eq:processes}
e^+ e^- & \to & f \bar{f}, \nonumber\\
\gamma & \to & f \bar{f}, \nonumber\\
e f &\to& e f.
\end{eqnarray}
The rates for these reactions are proportional to $\epsilon^2$ while
those for other conceivable processes such as $\gamma\gamma\to f
\bar{f}$ or $\gamma f\to\gamma f$ are proportional to $\epsilon^4$ and
thus much smaller.  At low temperatures, the electron-positron number
density is exponentially suppressed, making the $f\bar f$ production
rate approach zero quickly when $T\lsim m_e$.  
The third process in Eq.~(\ref{eq:processes})
does not contribute to the production of milli-charged particles but
helps to achieve kinetic equilibrium.

For a calculation of the $f \bar f$ production rates we use Boltzmann
statistics for all species. Then the rate for the pair process
is~\cite{KT90} $\Gamma_{e^+e^- \to f \bar{f}} = n_e \langle \sigma |v|
\rangle$, where $\langle \sigma |v| \rangle$ is the thermally averaged
annihilation cross section. It can be expressed as~\cite{GG91}
\begin{equation}
\langle\sigma|v|\rangle=\frac{1}{8 m_e^4 T K_2^2(m_e/T)}
\int_{4 m_e^2}^{\infty} ds\, s^{1/2} 
(s - 4 m_e^2) K_1(s^{1/2}/T)\,\sigma_{\rm CM},
\end{equation}
where $K_n$ is a modified Bessel function of the second kind and
$\sigma_{\rm CM}$ is the center-of-mass cross section.  The squared
matrix element is
\begin{eqnarray}
\frac{1}{4} \sum |M|^2&=&\frac{8 \epsilon^2 e^4}{(p_{e^-}+p_{e^+})^4} 
\left[(p_{e^-}p_f)(p_{e^+}p_{\bar{f}})
+ (p_{e^-}p_{\bar{f}})(p_{e^+}p_f) 
+ m_f^2 (p_{e^-}p_{e^+}) \right. \nonumber \\ 
&& \kern6em \left.
{} + m_e^2 (p_f p_{\bar{f}}) + 2 m_e^2 m_f^2\right],
\end{eqnarray}
which translates into the usual CM cross section $\sigma_{\rm CM}
=(4\pi/3)\,\epsilon^2 \alpha^2\,s^{-1}$ if all the particles can be
considered massless. In this limit we find
\begin{equation}
\langle\Gamma_{e^+e^- \to f \bar{f}}\rangle
= \frac{\zeta_3}{2 \pi}\, \epsilon^2 \alpha^2 T,
\end{equation}
where $\zeta$ refers to the Riemann zeta function. 

The decay rate of a transverse plasmon (photon) with energy $\omega$
and momentum $k$ into massless milli-charged particles is
$\Gamma_{\gamma\to f\bar f}=\epsilon^2\alpha Z (\omega^2 - k^2)/3
\omega$~\cite{GGR}.  In a relativistic plasma, the plasma frequency is
given by $\omega_{\rm P}^2=(4\pi/9)\alpha T^2$.  Except for the
low-energy part of the blackbody photon spectrum we have $\omega\gg
\omega_{\rm P}$, a limit where the dispersion relation is $\omega^2 -
k^2=\frac{3}{2} \omega_{\rm P}^2$ and the wave-function
renormalization factor is $Z=1$~\cite{GGR}. With Boltzmann statistics
we find $\langle \omega^{-1}\rangle=(2T)^{-1}$ so that finally the
average plasmon decay rate is
\begin{equation}
\langle \Gamma_{\gamma\to f\bar f}\rangle
=\frac{\pi}{9}\,\epsilon^2 \alpha^2 T.
\end{equation}
Comparing the rates for the two processes we find
\begin{equation}
\frac{\langle\Gamma_{e^+e^-\to f\bar f}\rangle}
{\langle\Gamma_{\gamma\to f\bar f}\rangle}
=\frac{9\zeta_3}{2\pi^2}=0.55,
\end{equation}
implying that the plasma process is actually somewhat more important
than pair annihilation. Notice, however, that the plasma process
is only important if $m \lsim \omega_P/2$. During BBN, the plasma
frequency is roughly given by $\omega_P \simeq \sqrt{(4\pi/9)\alpha} T
\simeq 0.1 \, T$, so that at $T=1$ MeV, where BBN begins,
the plasma process is only important for $m \lsim 50$ keV. 
Therefore the plasma process is important only for milli-charged particles
with masses in the keV region, where astrophysical bounds from
stellar evolution are much more stringent. Thus, it seems safe
no neglect the plasma process in the BBN calculations.

\subsection{Solving the Boltzmann Equation}

The standard procedure for calculating the evolution of number and
energy density of a given species is to use the Boltzmann collision
equation.  It describes the time evolution of the single particle
distribution function, $f_1$, of any given species. In the expanding
universe, formally this equation can be written in our case
as~\cite{KT90}
\begin{equation}
\frac{\partial f_1}{\partial t} - H p\, \frac{\partial f_1}
{\partial p}
= (C_{\rm ann} + C_{\rm el})\,[f_1],
\label{eq:boltz}
\end{equation}
where $H \equiv \dot{R}/R$ is the Hubble expansion parameter.  On the
right-hand side, $C_{\rm ann}$ is the collision operator describing
pair annihilation.
$C_{\rm el}$ is the elastic-scattering term which includes the
combined effect of scattering on electrons, positrons, and photons.

The $e^+e^-$ process is of the form $1+2 \to 3 + 4$ so that the
collision term can be written generically as~\cite{KT90}
\begin{equation}\label{integral}
C_{\rm ann}[f_1]  =  \frac{1}{2E_{1}}\int d^{3}\tilde{p}_{2}
d^{3}\tilde{p}_{3}d^{3}\tilde{p}_{4}
\Lambda(f_{1},f_{2},f_{3},f_{4})
\sum |M|^{2}\delta^{4}
({\it p}_{1}+{\it p}_{2}-{\it p}_{3}-{\it p}_{4})(2\pi)^{4}.
\end{equation}
Here, $d^{3}\tilde{p}\equiv d^{3}p/[(2 \pi)^{3} 2 E]$, $\sum |M|^{2}$
is the squared matrix element, and ${\it p}_{i}$ is the four-momentum
of particle $i$. The phase-space factor is $\Lambda \equiv
f_3 f_4 (1-f_1)(1-f_2)-f_1 f_2 (1-f_3)(1-f_4)$.

The dominant elastic scattering term from scattering on electrons and
positrons suffers from the usual infrared Coulomb divergence. It can
be regulated by including Debye screening effects in the medium.
However, instead of treating elastic scattering in any detail we
calculate the evolution of the milli-charged particle ensemble for two
extreme cases, a) No elastic scattering, and b) Elastic scattering is
assumed to be efficient enough to bring the milli-charged species into
complete kinetic equilibrium at all times.  It will turn out that the
difference between these two cases is quite small, justifying our
neglect of a detailed treatment of elastic scattering.

The dynamical evolution of the cosmic scale factor is governed by the
Friedmann equation and the equation of energy
conservation~\cite{KT90},
\begin{eqnarray}
H^2 & = & \frac{8 \pi G \rho}{3}, \\
\frac{d}{dt}(\rho R^{3}) & = & - P \frac{d}{dt}(R^{3}),
\end{eqnarray}
where $\rho$ is the energy density and $P$ the pressure, both
including the effect of milli-charged particles.

We have solved these equations to find the energy density of
milli-charged particles.  We have always neglected their mass, an
approximation which is accurate for $m_\epsilon$ up to the electron
mass.  For larger masses the milli-charged particles can pair
annihilate into electron-positron pairs at low temperatures.  We have
always taken a zero initial population of milli-charged particles, an
assumption which is valid if they interact only via their electric
charge.

Adding energy density to the cosmic plasma around the epoch of
$e^+e^-$ annihilation perturbs Big Bang Nucleosynthesis \cite{KT90}.
In order to calculate constraints on $\epsilon$ we have used the
nucleosynthesis code of Kawano \cite{K92}, modified to include the
energy density in milli-charged particles as well as the change in
neutrino-to-photon temperature ratio. The effect of the milli-charged
particles corresponds to additional neutrino degrees of freedom of
\begin{equation}\label{eq:neffofeps}
\Delta N_{\rm eff}=0.69\times10^{17}\,\epsilon^2
\times\cases{1&no elastic scattering,\cr
1.39&full equilibrium,\cr}
\end{equation}
an approximation is valid for $\epsilon \lsim 6 \times 10^{-9}$.
This is more than sufficient for our purposes since
all values of $\epsilon$ higher than this produce such a large 
$\Delta N_{\rm eff}$ that they are excluded by a huge margin.
As expected, the
effect of elastic scattering on the energy density in milli-charged
particles is quite small as this process does not produce additional
particles.

\subsection{BBN Limits}

There are still some unresolved issues regarding the observationally
determined values of the primordial light-element abundances. For the
past few years there have been two favoured solutions, namely the
so-called High-Helium/Low-Deuterium and the Low-Helium/High-Deuterium
solutions~\cite{BBN}.  There seems to be growing consensus that the
first is the correct one.  Nevertheless any bound derived from
nucleosynthesis should be used with some caution since the question of
primordial deuterium and helium abundances is not yet fully settled.

In the present paper we shall use the data on primordial helium
obtained by Izotov and Thuan~\cite{IT98} and the deuterium data from
Burles and Tytler~\cite{BT98}
\begin{eqnarray}
Y_P & = & 0.244 \pm 0.002, \\
D/H & = & (3.39 \pm 0.25) \times 10^{-5},
\end{eqnarray}
where errors are estimated $1\sigma$ uncertainties.
These data give the High-He/Low-D solution and are completely
consistent with standard BBN for a baryon-to-photon ratio of $\eta =
(5.1 \pm 0.3) \times 10^{-10}$ \cite{BT98}.
Used together, these data provide the constraint
\begin{equation}
N_{\rm eff} = 2.98 \pm 0.33 \,\,\, (90\% \,\, {\rm C.L.}),
\end{equation}
on the effective number of neutrinos,
thus tightly constraining any non-standard nucleosynthesis scenario.

Together with the less restrictive case (no elastic scattering) of 
Eq.~(\ref{eq:neffofeps}) BBN then implies a limit 
\begin{equation}
\epsilon < 2.1 \times 10^{-9}.
\end{equation}
This bound applies to masses in the regime $m_\epsilon \lsim m_e$.
While our result is very similar to what one finds from simple
dimensional estimates, our calculation is quantitative and the errors
are controlled.


\section{Stellar Evolution}

\subsection{Globular Clusters}

New low-mass particles will be produced in the hot and dense medium in
the interior of stars and subsequently escape. This new energy-loss
channel leads to observational modifications of the standard course of
stellar evolution and thus can be used to set limits on the particle's
interaction strength. For milli-charged particles, limits have been set
from red giants \cite{neu,D+I,DCB,D+P,M+N,HRW94}, horizontal-branch
(HB) stars~\cite{GGR}, white dwarfs~\cite{D+I,DCB}, and supernova (SN)
1987A \cite{M+R}.  For small masses of the milli-charged particles, the
most restrictive limits arise from HB stars and low-mass red giants in
globular clusters.

At the end of the main-sequence evolution of normal stars, the
hydrogen in the inner part has been consumed, leaving the star with a
core consisting mainly of helium. In low-mass stars, this helium core
reaches degeneracy before it is hot and dense enough to be ignited.
This process is very dependent on density and temperature, and even
minor changes in these quantities produce observable changes in the
brightness at the tip of the red-giant branch in globular clusters.
Therefore, the core mass at helium ignition as implied by the
color-magnitude diagrams of several globular clusters implies a limit
on any new energy-loss channel.  After helium ignition, the stars move
to the horizontal branch where they burn helium in their core. A new
energy-loss mechanism will lead to an accelerated consumption of
nuclear fuel, shortening the helium-burning lifetime which can be
``measured'' by number counts of HB stars in globular clusters.  When
applied to a new energy-loss channel, usually one of these arguments
is more restrictive~\cite{GGR}.  For example, a putative neutrino
dipole moment will add to the efficiency of the plasmon decay process
$\gamma\to\nu\bar\nu$ and thus enhance neutrino losses. The
helium-ignition argument provides far more restrictive limits than the
helium-burning lifetime argument because the plasmon decay is more
effective in the degenerate red-giant core. On the other hand, axion
losses by the Primakoff process are more effective in the
nondegenerate cores of HB stars so that the helium-burning lifetime
argument yields more restrictive limits.

The emission of milli-charged particles is a special case in that both
arguments yield comparable limits of about~\cite{GGR}
\begin{equation}
\epsilon \leq 2 \times 10^{-14}.
\end{equation}
The reason is that the rate of the plasma decay process $\gamma\to
f\bar f$ for milli-charged particles is proportional to $\omega_{\rm
  P}^2$, as opposed to the magnetic-dipole case ($\omega_{\rm P}^4$)
or standard-model neutrino case ($\omega_{\rm P}^6$). The low power of
the plasma frequency implies that the emission rate per unit mass is
almost independent of density.  This, in turn, implies that the core
expansion caused by helium ignition leaves the energy-loss rate per
unit mass nearly unchanged, while it is ``switched off'' for the
dipole-moment case, or ``switched on'' for the axion case.

An average value for the plasma frequency in the core of a
globular-cluster star before helium ignition is $\omega_{\rm P} \simeq
10~{\rm keV}$ (in the center it is about twice that) while it is about
2~keV in the core of a HB star. Therefore, the helium-ignition
argument constrains milli-charged particles with masses up to about
5~keV.

Of course, this sort of argument applies only if the interaction
strength is small enough that the milli-charged particles escape freely
once produced in the stellar core; one can check that this is the case
for $\epsilon\lsim 10^{-8}$~\cite{GGR}.  For larger charges, the
particles would contribute to the transfer of energy rather than
carrying away energy directly.  In order to avoid observable
consequences, the efficiency of energy transfer would have to be less
than that of photons, i.e.\ the mean free path would have to be very
short. It is unlikely that there exists an allowed range of
milli-charges $\epsilon\lsim 1$ where all stars would be left
unchanged. However, since laboratory limits and the BBN argument
exclude values for $\epsilon$ above $10^{-8}$ anyway, a detailed
discussion of the trapping limit is not warranted.

\subsection{White Dwarfs}
 
The observed population of hot young white dwarfs is consistent with
cooling by surface emission of photons and by volume emission of
neutrinos produced by plasmon decay via Standard Model
interactions~\cite{BDB,GGR}.  Blinnikov and Dunina-Barkovskaya
\cite{BDB} set a bound on the neutrino magnetic moment, $\mu_{\nu} <
10^{-11} \mu_B$ (where $\mu_B = e/2 m_e$), by requiring that the
additional neutrino emission not cool the white dwarfs faster than
observed. We can translate this bound into one on milli-charged
particles.

For neutrinos or neutrino-like particles coupling to the photon via a
dipole moment, the energy loss rate can be written as
\begin{equation}
Q_\mu \propto 
\frac{\mu^2}{2} \left(\frac{\omega_{\rm P}^2}{4 \pi} \right)^2 \, Q_2,
\end{equation}
whereas for milli-charged particles, it is given by 
\begin{equation}
Q_\epsilon \propto \epsilon^2 \alpha\,\frac{\omega_{\rm P}^2}{4\pi} \, Q_1,
\end{equation}
where $Q_1/Q_2$ is a factor of order unity~\cite{GGR}. 
Here, we set $Q_1 = Q_2 = 1$.
One then obtains
\begin{equation}
\frac{Q_\epsilon}{Q_\mu} = 2.09\, \frac{\epsilon_{14}^2}{\mu_{12}^2}
\left(\frac{\omega_{\rm P}}{10~{\rm keV}}\right)^{-2},
\end{equation}
where $\epsilon_{14}\equiv10^{14}\,\epsilon$ and
$\mu_{12}\equiv 10^{12}\,\mu/\mu_B$.
In order to calculate an average value of this ratio over the
entire star, we need to perform an ``emissivity-average'' of the form
\begin{equation}
\left\langle \frac{Q_\epsilon}{Q_\mu}\right\rangle 
=\frac{\int dr r^2 
\frac{Q_\epsilon}{Q_\mu} Q_\mu}{\int dr r^2  Q_\mu}.
\end{equation}
In order to do these integrals, we assume that the white dwarf is
a polytrope of index $n=3/2$, so that
\begin{equation}
\left\langle \frac{Q_\epsilon}{Q_\mu}\right\rangle 
=2.40 \times 10^{5} \rho_{c,6}^{-1}
\frac{\epsilon_{14}^2}{\mu_{12}^2}
\frac{\int d \xi \xi^2 \theta^n}{\int d \xi \xi^2 \theta^{2n}},
\end{equation}
where $r = \alpha \xi$ and $\rho = \rho_c \theta^n$ \cite{ST83}.
For a white-dwarf mass of 0.7 solar masses, this gives
\begin{equation}
\left\langle \frac{Q_\epsilon}{Q_\mu} \right\rangle = 
0.34\,\frac{\epsilon_{14}^2}{\mu_{12}^2}.
\end{equation}
If we then demand that the emission rate due to milli-charges is not
larger than for neutrino dipole moments, we find
\begin{equation}
\epsilon_{14} \leq 17.
\end{equation}
This bound is consistent with that found in Ref.~\cite{D+I}.

One may worry that the way the emissivity-average is performed has
consequences for the bound. One could instead use
\begin{equation}
\left\langle \frac{Q_\epsilon}{Q_\mu}\right\rangle 
= \frac{\int dr r^2 
\frac{Q_\epsilon}{Q_\mu} Q_\epsilon}{\int dr r^2  Q_\epsilon}.
\end{equation}
This yields $\epsilon_{14} \leq 15$ so that the way the average is
performed matters little for the final result.

The emissivity-averaged plasma frequency is $\langle(10~{\rm
  keV}/\omega_{\rm P})^{2}\rangle = 0.16$ so that $\langle \omega_{\rm
  P} \rangle = 25~{\rm keV}$.  Therefore, the white-dwarf bound
applies to milli-charged particles with $m_\epsilon \lsim 10$ keV,
similar to the red-giant case. 

\subsection{Supernova 1987A}

Finally, the stellar energy-loss argument can be applied to SN~1987A
where the number of neutrinos detected at Earth agree roughly with
theoretical expectations. If there are other particles contributing to
the cooling of the proto neutron star, this will reduce the neutrino
fluxes and the duration of the neutrino signal.  Therefore, if we
assume that such hypothetical particles freely stream from the core
where they are produced, one can put an approximate bound on the
allowed loss rate \cite{GGR} of
\begin{equation}
\langle Q/\rho\rangle
\lsim 10^{19}~{\rm erg~g^{-1}~s^{-1}},
\end{equation}
to be calculated at average core conditions of about
$3\times10^{14}~{\rm g~cm^{-3}}$ and 30~MeV for density and
temperature, respectively. $ 10^{19}~{\rm erg~g^{-1}~s^{-1}}$
is the  proto-neutron
star's average rate of energy loss  to neutrinos.

The main production process for milli-charged particles in a SN core is
the plasmon process. In a degenerate, relativistic electron plasma it
is orders of magnitude larger than $e^+e^-$ annihilation. The
energy-loss rate is \cite{GGR}
\begin{equation}
Q_{\rm P} = \frac{8\zeta_3}{9 \pi^3}\, \epsilon^2 \alpha^2 
\left(\mu_e^2 + \frac{\pi^2 T^2}{3}\right) T^3\,
Q_1,
\end{equation}
where $Q_1$ is again a factor of order unity. Setting
$Q_1=1$, one finds
\begin{equation}
\epsilon \leq 1 \times 10^{-9}
\end{equation}
This bound matches the one found by Mohapatra and
Rothstein~\cite{M+R}.  However, they considered nucleon-nucleon
bremsstrahlung as a production process and notably an amplitude where
the electromagnetic current is coupled to an intermediate charged
pion. We believe that a naive perturbative calculation of this
process in a nuclear medium can be unreliable \cite{GGR}.
However, since the bounds are so similar, a detailed study of the
nucleon process is not warranted.

If $\epsilon$ is much larger than our limit, the milli-charged
particles are trapped inside the proto neutron star, and can only
escape via diffusion. The main process which keeps milli-charged
particles trapped is Coulomb scattering on protons
\footnote{The scattering on electrons was
instead considered in \cite{M+R}, but
this is less important than protons.}. The differential scattering
cross section on nonrelativistic protons is
\begin{equation}
\frac{d\sigma}{d\Omega}=2\epsilon^2\alpha^2\,
\frac{E^2(1+\cos\theta)}{|{\bf q}|^4}
\end{equation}
where ${\bf q}$, with $|{\bf q}|^2=2E^2(1-\cos\theta)$, is the momentum
transfer, $E$ the milli-charged particle's energy, and $\theta$ the
scattering angle. This cross section is strongly forward peaked and
thus not a good measure for particle trapping: a particle which is
deflected by a small angle continues its way out of the star
essentially as if it had not been scattered at all. Therefore, we
rather consider the usual {\it transport\/} cross section which
includes an additional weight $(1-\cos\theta)$. Therefore,
\begin{equation}
\frac{d\sigma_{\rm T}}{d\Omega}=\epsilon^2\alpha^2\,
\frac{(1+\cos\theta)}{{\bf q}^2}
\end{equation}
is a more adequate measure for the effectiveness of particle trapping.

In addition, we need to include proton-proton correlations induced by
their mutual Coulomb repulsion, i.e.\ we need to include screening
effects. This is achieved by multiplying the cross section with the
static structure function $S({\bf q})={\bf q}^2/({\bf
  q}^2+k_{\rm S}^2)$ so that finally
\begin{equation}
\frac{d\sigma_{\rm T,eff}}{d\Omega}=\epsilon^2\alpha^2\,
\frac{(1+\cos\theta)}{{\bf q}^2+k_{\rm S}^2}.
\end{equation}
For nonrelativistic, nondegenerate protons, the screening scale
$k^2_{\rm S}$ is given by the proton Debye scale $k^2_{\rm S}=k^2_{\rm
  D}=4\pi\alpha n_p/T$ with $n_p$ the proton density. It would have
been incorrect to use the electron screening scale since the
background of degenerate electrons is much ``stiffer'' than the
protons, i.e.\ most of the polarization of the plasma by a test charge
is due to the protons. With these modifications we find a total cross
section
\begin{equation}\label{eq:cross}
\sigma_{\rm T,eff}=\frac{2\pi\epsilon^2\alpha^2}{E^2}\,
\left[\frac{(2+z)}{2}\ln\left(\frac{2+z}{z}\right)-1\right],
\end{equation}
where
\begin{equation}
z\equiv\frac{k_{\rm D}^2}{2E^2}=\frac{2\alpha}{3\pi}\,\eta_e^3
\left(\frac{T}{E}\right)^2
=0.335\,\left(\frac{\eta_e}{6}\right)^3\left(\frac{T}{E}\right)^2
\end{equation}
and $\eta_e=\mu_e/T$ is the electron degeneracy parameter. We have
used that $n_p=n_e=\mu_e^3/3\pi^2$. The transport mean free path is
then found to be
\begin{equation}\label{eq:mfp}
\lambda^{-1}_{\rm T,eff}\equiv
\sigma_{\rm T,eff}n_p=\epsilon^2\alpha T\,z
\left[\frac{(2+z)}{2}\ln\left(\frac{2+z}{z}\right)-1\right],
\end{equation}
where we have expressed $n_p$ in terms of $k_{\rm D}$.  Taking
$T=30~{\rm MeV}$ as a typical value, and using 1 for the $z$-dependent
expression, we find that $\lambda_{\rm T,eff}$ exceeds the
proto-neutron star radius of about 10~km for
$\epsilon\lsim 1 \times10^{-8}$.

When $\epsilon$ is larger than this, the particles no longer freely
escape. Rather, a ``photo-sphere'' for the milli-charged particles 
is created. For them to be less effective at carrying away energy,
their transport cross section must be about as large as that for
neutrinos or larger. Very crudely, we must compare
Eq.~(\ref{eq:cross}) with the weak-interaction transport cross 
section for neutral-current scattering on nucleons,
\begin{equation}
\sigma_{\rm T,weak}=\frac{2G_F^2E^2}{3\pi}(C_V^2+5C_A^2),
\end{equation}
where $|C_A|\approx1.26/2$ for protons and neutrons, while
$|C_V|\approx0$ for protons and $1/2$ for neutrons. If we use
$z=0.01$ near the neutrino sphere, the expression in square brackets
in Eq.~(\ref{eq:cross}) is about 4. With this value we find
\begin{equation}
\left(\frac{\sigma_{\rm T,eff}}{\sigma_{\rm T,weak}}\right)^{1/2}
\approx
1.2\times10^{7}\,\epsilon\,\left(\frac{20~{\rm MeV}}{E}\right)^2.
\end{equation}
Taking an average energy of 20~MeV we conclude that $\epsilon$ should
exceed approximately  $8\times10^{-8}$ for emission of
milli-charged particles to be less important than neutrinos.

While this derivation is somewhat crude, we conclude that milli-charged
particles in the range $10^{-9}\lsim\epsilon\lsim 10^{-7}$ are
excluded. The plasma frequency in a proto-neutron star is roughly
10~MeV so that this bound applies to milli-charged particles with mass
below about 5~MeV. The upper bound on $\epsilon$ will also apply
to the model with a second $U(1)$. The
trapping limit $\epsilon > 10^{-7}$  could be slightly
modified by the presence of the paraphoton,
but this area of parameter space is already 
ruled out by nucleosynthesis, so we do
not discuss this further.
 In summary, SN~1987A excludes only a very narrow
sliver of parameter space in addition to what is excluded by other
arguments (Fig.~1).


\section{Milli-Charged Neutrinos}

In the Standard Model with massless neutrinos, there are four
anomaly-free $U(1)$ symmetries, corresponding to the Standard Model
hypercharge $Y_{\rm SM}$, and $\{ B/3 - L_i \}$.  $B$ is baryon
number, and $L_i$ are the lepton numbers of the three lepton families.
The hypercharge operator can be redefined to be \beq Y' = Y_{\rm SM} +
2 \sum_i \epsilon_i \left( \frac{B}{3} - L_i \right)
\label{Y'}
\eeq without making the theory anomalous. This gives electric charge
$\epsilon_i$ to $\nu_i$ and generates a proton-electron charge
difference $\epsilon_e$.  Constraints on this model and variants have
been discussed in \cite{FJL,TT1,B+V}.

It is more difficult to give charge to massive neutrinos in this way,
as discussed in \cite{BM2}. The solar and atmospheric neutrino
deficits can be explained by masses, mixing the three Standard Model
neutrinos. If this is the case, lepton number and the lepton
flavours are not conserved, so the neutrinos cannot acquire electric
charge by the redefinition of equation (\ref{Y'}). If there are three
additional gauge singlet ``right-handed'' neutrinos {\it without}
Majorana masses, the neutrino masses can conserve $B-L$ which would
allow the three flavours to have the same charge $\epsilon$. 
In this case, the
observed neutrality of matter \cite{MM} implies $\epsilon < 10^{-21}$
\cite{FJL}. It would be possible for $\nu_{\mu}$ and $\nu_{\tau}$ to
have significantly larger charges than this if the observed solar and
atmospheric neutrino deficits are not due to a neutrino mass matrix
mixing the three Standard Model neutrinos\footnote{For instance, it
  has been suggested that the solar neutrino deficit could be caused
  by the deflection of milli-charged massless neutrinos in the Sun's
  magnetic field \cite{IJ}}.  Bounds on the electric charge of
$\nu_{\tau}$ were calculated in \cite{FL} without assuming any
relation between the tau neutrino charge and the electric charge of
the other two neutrinos.  We do not consider this possibility.


\section{Conclusion}

We have updated the bounds from astrophysics, cosmology
and laboratory experiments on fermions with
electric charge $\epsilon e$, where $\epsilon < 1$. These
``milli-charged'' particles could be neutrinos or new
fermions from beyond-the-Standard-Model. 
If milli-charged neutrinos have mass, and the Standard Model
particle content is non-anomalous, the electric charge of neutrinos
is constrained to be $ < 10^{-21}$. The
updated bounds on milli-charged particles from beyond-the-Standard-Model 
are presented in figure 1, for both the cases where there is, and there
is not, a paraphoton. For masses $< 5$ keV, a
fermion with electric charge $ 2 \times 10^{-14} < \epsilon e < 1$ is
ruled out.  An electric charge $  \epsilon > 10^{-8}$ is
ruled out up to masses $\sim$ MeV. 
Milli-charged particles  could
be possible for masses between an MeV and a TeV.


\acknowledgments

We are grateful to Andrzej Czarnecki and Alick Macpherson for
informative discussions.  In Munich, this work was supported, in part,
by the Deutsche Forschungsgemeinschaft under grant No.\ SFB-375.  In
Copenhagen, it was supported by a grant from the Carlsberg Foundation.


\end{document}